\documentclass[bibyear]{aa} 
\usepackage{amsmath, amssymb}
\usepackage{subfig}
\usepackage{booktabs}
\usepackage{natbib}
\usepackage{url}
\usepackage{hyperref}
\usepackage{longtable}
\usepackage{tablefootnote}
\usepackage{graphicx}
\defcitealias{BaglioZelati2023}{Baglio \& Coti Zelati et al., (2023)}

\usepackage{txfonts}

\begin{document}

   \title{High-temporal-resolution optical spectroscopic observations of the transitional millisecond pulsar PSR J1023+0038}

   \author{M. M. Messa
          \inst{1}\inst{,2}, P. D'Avanzo\inst{2}, F.~Coti~Zelati\inst{3,4,2}, M. C. Baglio\inst{2}, S. Campana\inst{2}
          }

   \institute{
   Università degli Studi di Milano, Dipartimento di Fisica, Via Celoria 16, 20133 Milano (MI), Italy;\\
            \email{marco.messa@unimi.it}
         \and
             INAF, Osservatorio Astronomico di Brera, Via E. Bianchi 46, 23807 Merate (LC), Italy
        \and
        Institute of Space Sciences (ICE, CSIC), Campus UAB, Carrer de Can Magrans s/n, E-08193, Barcelona, Spain
        \and
        Institut d'Estudis Espacials de Catalunya (IEEC), Carrer Gran Capit\`a 2--4, E-08034 Barcelona, Spain
             }
    
   \date{Received month day, year; accepted month day, year}

\titlerunning{PSR J1023}
\authorrunning{M. M. Messa et al.}
 
  \abstract
  {
Transitional millisecond pulsars (tMSPs) represent a dynamic category of celestial sources that establish a crucial connection between low-mass X-ray binaries and millisecond radio pulsars. These systems exhibit transitions from rotation-powered states to accretion-powered ones and vice versa, highlighting the tight evolutionary link expected by the so-called recycling scenario. In their active phase, these sources manifest two distinct emission modes named high and low, occasionally punctuated by sporadic flares. Here, we present high-time-resolution spectroscopic observations of the binary tMSP J1023+0038, in the sub-luminous disc state. This is the first short-timescale ($\sim$ 1 min) optical spectroscopic campaign ever conducted on a tMSP. The campaign was carried out over the night of June 10, 2021 using the Gran Telescopio Canarias. The optical continuum shows erratic variability, without clear evidence of high and low modes or of orbital modulation. Besides, the analysis of these high-temporal-cadence spectroscopic observations reveals, for the first time, evidence for a significant (up to a factor of $\sim 2$) variability in the emission line properties (equivalent width and full width half maximum) over a timescale of minutes. Intriguingly, the variability episodes observed in the optical continuum and in the emission line properties seem uncorrelated, making their origin unclear.} 


\maketitle
%

\section{Introduction}
Millisecond radio pulsars (MSPs) are rapidly rotating magnetic neutron stars (NSs) whose fast spin originates through the transfer of matter in a low-mass X-ray binary (LMXB, \citealt{Alpar1982}; \citealt{Radhakrishnam&Srinivasan}). The discovery of the first 'transitional' MSP (tMSP), PSR J1023+0038 (hereafter J1023; \citealt{Archibald2009}), established a connection between rotation-powered MSPs and accretion-powered NSs in LMXBs. These systems are observed to switch back and forth between a rotation-powered radio pulsar state ('pulsar state') and a state characterised by X-ray pulsations and accretion disc features in the optical spectra ('disc state'). Besides J1023, three other tMSPs systems are known so far: J18245-2452 (M28I; \citealt{Papitto2013}), XSS J12270–4859 (\citealt{Bassa2014}; \citealt{DeMartino2013}), plus a few candidates: RXS J154439.4-112820 (\citealt{Bogdanov2015+}; \citealt{Bogdanov2016}; \citealt{Britt2017}), CXOU J110926.4-650224 (\citealt{CotiZelati2019}), 4FGL J0407.7–5702 (\citealt{Miller2020}; \citealt{Kennedy2020}), 3FGL J0427.9-6704 (\citealt{Strader2016}; \citealt{Li2020}), Terzan 5 CX10 (\citealt{Bahramian2018}) and XMM J174457-2850.3 (\citealt{Degenaar2014}; \citealt{Degenaar2015}).\\ 
J1023 was discovered by \cite{Bond2003} as part of the Faint Images of the Radio Sky at Twenty Centimeters’ (FIRST) survey. In 2001, it was categorised as a cataclysmic variable because the optical counterpart to the radio source exhibited rapid, short-duration flickering and a blue optical spectrum with double-peaked emission lines, indicative of an accretion disc (\citealt{Szkody2003}). However, optical photometry conducted in 2003 (\citealt{Woudt2004}) and 2004 (\citealt{Homer2006}) did not show the intense, rapid flickering events observed in the 2001 light curve, indicating a change in the system's state. Only 4.75-hour single-humped modulation was detected (\citealt{Woudt2004}). Indeed, \cite{Thorstensen_2005} confirmed this state transition, as the optical spectrum taken in 2003 displayed strong absorption features and lacked the prominent emission lines seen in the 2001 spectrum. They conducted a time-resolved optical spectroscopic and photometric study of J1023. They identified the companion star as a late-type G5 star with an absorption-line radial velocity semi-amplitude of 268 ± 3 km s$^{-1}$ modulated at the 4.75-hr orbital period. Optical light curves taken in 2004 revealed a single-humped modulation, attributed to the companion star strongly irradiated by the pulsar particle wind. The combination of photometric and radial velocity studies led to the conclusion that the system was not a cataclysmic variable but, rather, an X-ray binary hosting an NS. This X-ray binary scenario also explained the 2004 X-ray observations, which were characterised by a dominant hard X-ray power-law component (\citealt{Homer2006}). The discovery of a 1.69-ms radio pulsar in a 4.75-hr binary system (\citealt{Archibald2009}) made J1023 the first system showing the potential to alternate between an X-ray state and a radio pulsar phase powered by rotation.\\
In 2013, J1023 underwent a sudden and significant increase in the emission levels at both X-ray and gamma-ray frequencies, amplified by a factor of 5–10. Simultaneously, there was a noticeable enhancement in the emission at ultraviolet (UV) and optical frequencies, with a magnitude increase of 1–2. This notable shift in emissions coincided with the vanishing of the pulsed radio signal (\citealt{Stappers2014} and \citealt{Patruno2014}). Shortly after these events, double-peaked optical emission lines were detected, which provided further evidence for the formation of an accretion disc, as was documented by \cite{CotiZelati2014}. J1023 has since remained in this active state, maintaining an X-ray luminosity of $L_{X}\sim$ 7 $\times$ 10$^{33}$ erg s$^{-1}$ in the energy range of 0.3–80 keV (\citealt{CotiZelati2018}), based on a distance estimate of 1.37 kpc provided by \cite{Deller2012}. This phase is also called sub-luminous because it differs in brightness from the typical luminosity of NSs in their X-ray accretion state of the order of $L_{X} \sim $ 10$^{36}$-10$^{38}$ erg s$^{-1}$. During the disc state, the X-ray emission from J1023 exhibits a distinctive pattern of switching between two intensity modes, which have been labelled as high and low (\citealt{Papitto2013}). These transitions are occasionally punctuated by sporadic flares (see e.g., \citealt{Linares2014}; \citealt{Bogdanov2015}). The high mode in the X-ray band is predominant, occurring roughly 70\% of the time with a luminosity of $L_{X}\sim$ 7 $\times$ 10$^{33}$ erg s$^{-1}$, while the low mode manifests 20\% of the time with a luminosity of $L_{X}\sim$ 3 $\times$ 10$^{33}$ erg s$^{-1}$. Typically, the low mode persists for durations ranging from a few tens of seconds to a few minutes. The rate of change in intensity during these mode switches happens on timescales of $\approx$ 10 s. 
In addition, in the X-ray high-intensity mode, there are concurrent pulsations at X-ray, UV, and optical frequencies, which cease during the low mode (\citealt{Archibald2015}; \citealt{Papitto2019}; \citealt{Zanon2022}, and \citealt{Illiano2023}). The presence of optical pulsations at the spin period of J1023 discovered by \cite{Ambrosino2017} in the disc state cannot be explained by an accretion of matter onto the NS. This finding is interpreted as a strong indication that a rotation-powered pulsar is active in the system, even when the accretion disc is present (\citealt{Papitto2019}).\\ 
The pulsar generates a wind of particles that interacts with the innermost accretion flow generating a region of shock very close to the pulsar (within two times the light cylinder radius; \citealt{Papitto2019}), maintaining coherence and giving rise to the pulsated emission observed from optical up to X-ray frequencies (\citealt{Papitto2019}; \citealt{Veledina2019}). During low modes, it has been proposed that the shock moves outwards (\citealt{Papitto2019}) or that the disc penetrates the light cylinder radius and the system enters the propeller regime (\citealt{Veledina2019}). More recently, \citetalias{BaglioZelati2023} have shown that the high-to-low mode transition could be interpreted as being due to the evolution of the innermost accretion disc region into a compact radio jet, with the additional emission of discrete ejecta, and that the low-to-high mode switch is due to the re-enshrouding of the pulsar.\\
The emission at UV, optical, and near-infrared (NIR) frequencies is primarily influenced by the presence of the accretion disc and the irradiated companion star. In addition, occasional flaring and flickering are observed (\citealt{Shabaz2018}; \citealt{Kennedy2018}; \citealt{Papitto2018}; \citealt{Hakala2018}, \citealt{BaglioVincetelli2019}), which were interpreted as being due to a combination of reprocessing of the optical emission and direct NIR emission from plasmoids in the accretion flow that are channelled onto the NS and then expelled from the magnetosphere (\citealt{Shabaz2018}). Time-resolved spectroscopic studies carried out in the optical band were presented by \cite{Hakala2018}, who observed changes in the disc emission structure (specifically in the properties of the H{\fontsize{3mm}{4mm}\selectfont $\alpha$} emission line) using a single exposure time of 400 s and a 2.5 m telescope. These observations, together with Doppler tomography, suggested that a propeller effect might occur during flaring, as is suggested by \cite{Shahbaz2015}. Additional minor contributions to the overall emission are provided by the shock between the relativistic pulsar wind and the surrounding material and by a jet  (\citealt{CotiZelati2014}; \citealt{BaglioVincetelli2019}; \citetalias{BaglioZelati2023}), seen at X-rays, NIR, and optical wavelengths, respectively.\\
In this paper, we report the results of a high-temporal-cadence spectroscopic study of J1023 carried out by obtaining a spectrum in the optical band when the source was in its disc state every $\sim 50$ s with the 10 m Gran TeCan Telescope. The paper is organised as follows: in Sec. \ref{Sec::Data&analysis}, we describe our observations and the data reductions; the results are presented and discussed in Sec. \ref{Sec::Results} and \ref{Sec::Discussion}; and our conclusions are presented in Sec. \ref{Sec::Conclusion}. Throughout the paper, errors are at a 68\% confidence level unless stated otherwise.

\section{Data reduction and analysis} \label{Sec::Data&analysis}

J1023 spectra were taken on June 10, 2021 with the 10.4m Gran Telescopio Canarias (GTC) at La Palma. We used the OSIRIS (Optical System for Imaging and low-Intermediate-Resolution Integrated Spectroscopy, \citealt{OSIRIS2000}) instrument for long-slit spectroscopy with the R1000B grism, achieving a resolution of $\sim$ 1000 and a dispersion value of 2.12 $\AA/$pix, with spectral coverage from 3630 to 7500 $\AA$. Observations were carried out over one night from 21:22 UT to 22:25 UT, covering 22\% of the total orbital period of J1023; namely, from phase 0.13 to 0.35, based on the precise ephemerides of \cite{Zanon2022}\footnote{Here, phase 0 is defined as the superior conjunction of the pulsar (when the NS is behind the companion).}. Each spectrum had an exposure time of 20 s and overall we obtained 87 spectra, each every $\sim$ 50 s. This is the first time that such a high-cadence spectroscopic study has been performed on a tMSP (and on an NS LMXB in general). The data were reduced using the standard ESO-MIDAS procedures for bias subtraction, flat-field correction, and cosmic ray removal. All spectra were sky-subtracted and corrected for atmospheric extinction. Helium-argon lamp spectra were obtained for wavelength calibration during daytime and with the telescope vertically parked. The wavelength scale was then derived through third-order polynomial fits to 26 lines, resulting in an rms scatter of <0.06 Å. Instrumental flexures during our observations were then accounted for using atmospheric emission lines in the sky spectra. Finally, each spectrum was flux-calibrated through the observation of a standard spectrophotometric star during the same night.

\section{Results}\label{Sec::Results}

\subsection{Average spectrum}
\begin{figure}[h!]
\centering
\includegraphics[width=1\linewidth]{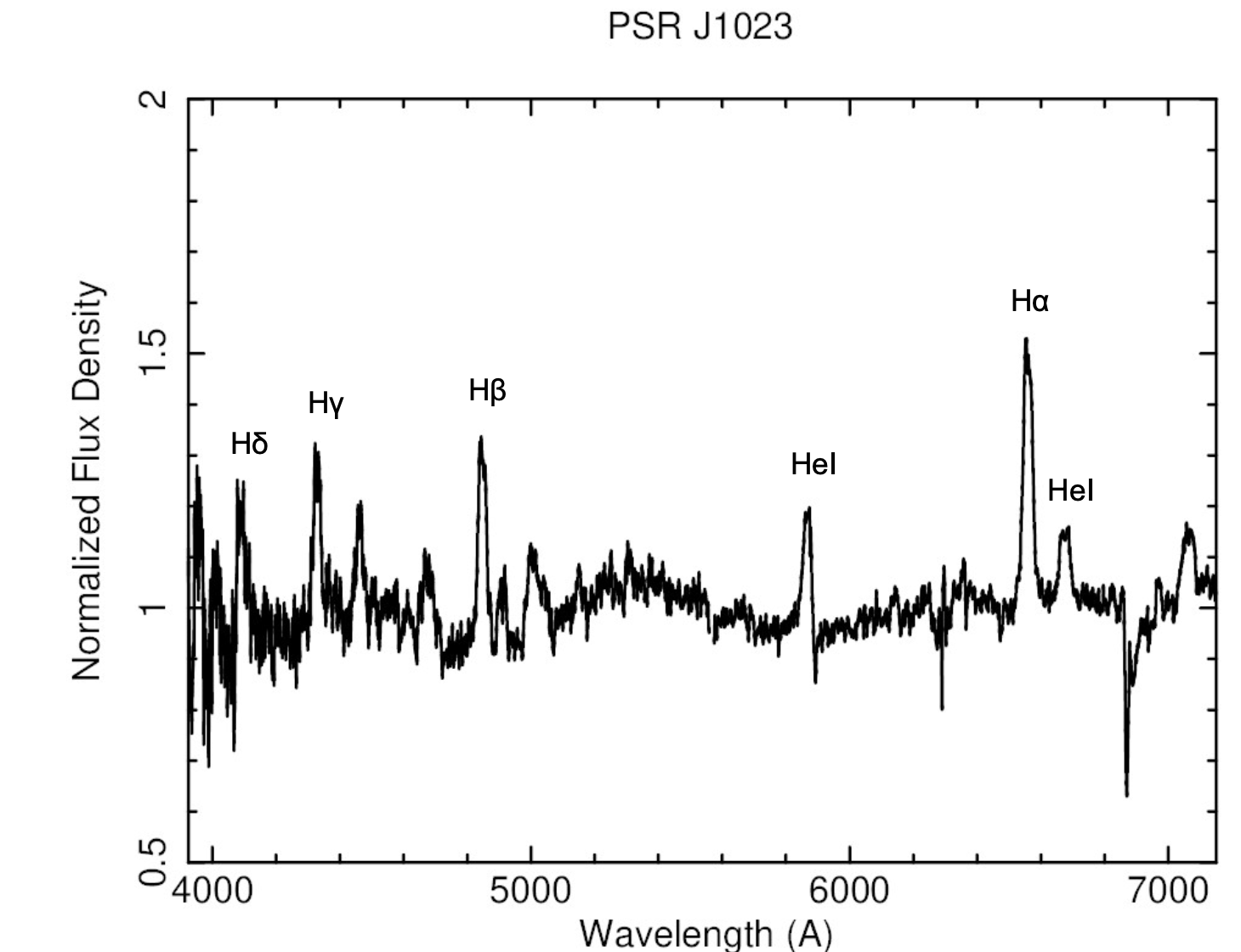}
\caption{Average spectrum of J1023 normalised to the emission of the continuum. The emission lines studied in this work are highlighted.
\label{im:spettro}}
\end{figure}

Fig. \ref{im:spettro} shows the normalised and averaged disc-state spectrum of J1023. We clearly detect H{\fontsize{3mm}{4mm}\selectfont $\alpha$}, H{\fontsize{3mm}{4mm}\selectfont $\beta$}, H{\fontsize{3mm}{4mm}\selectfont $\gamma$}, H{\fontsize{3mm}{4mm}\selectfont $\delta$}, He{\fontsize{3mm}{4mm}\selectfont I} ($\lambda \lambda$4472, 4921, 5016–5048, 5876, 6678, 7065 $\AA$), and He{\fontsize{3mm}{4mm}\selectfont II} at 4686 $\AA$. As has been shown in other work in the literature (\citealt{CotiZelati2014}), these lines display a double-horned profile that indicates the presence of an accretion disc.

\noindent

\subsection{Continuum emission}
\begin{figure}[h!]
\centering
\includegraphics[width=1.05\linewidth]{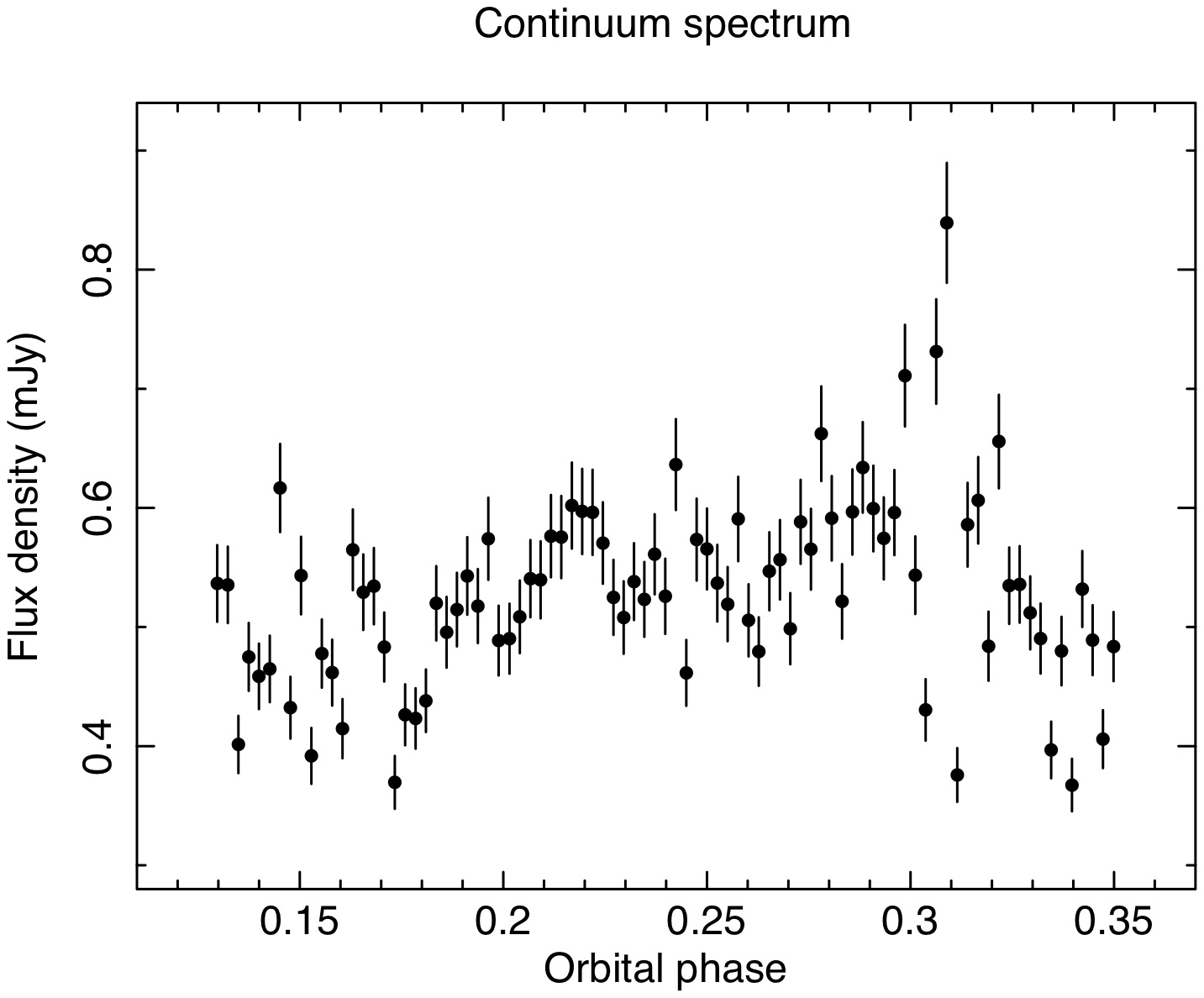}
\caption{Plot of the continuum (taken between 5200 and 5600 \AA) V-band flux as a function of the orbital phase. For each spectrum, the uncertainty of the flux density is given by the square root of the number of counts. 
\label{im:bandaV}}
\end{figure}

The initial focus was to analyse how the various spectral characteristics evolved over time. We started by examining the trend of the optical continuum, which serves as a proxy for the average brightness of the source in the optical range. We used the \texttt{molly}\footnote{\url{https://cygnus.astro.warwick.ac.uk/phsaap/software/molly/html/INDEX.html}} program to carry out this analysis, which allowed us to measure the variability of the continuum by choosing a wavelength range roughly centred in the V-band spectrum region, spanning from approximately 5200 to 5600 \AA. As can be seen from Fig. \ref{im:bandaV}, the optical continuum shows an erratic variability with possible flaring episodes at about phase 0.3. Despite the limited orbital coverage, no clear orbital modulation (sinusoidal or ellipsoidal) in the flux density seems to be present in our data. The analysis was also repeated for longer wavelengths (5900 to 6100 $\AA$), obtaining the same kind of variability.

\noindent

\begin{figure}[h!]
\centering
\includegraphics[width=0.90\linewidth]{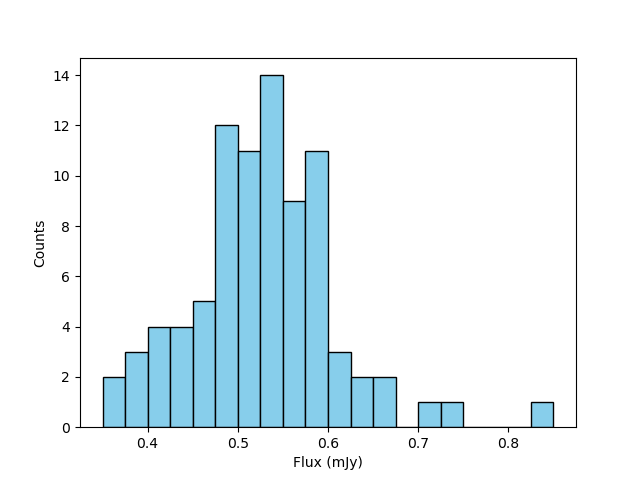}
\caption{Count distribution for the continuum V-band flux of J1023.
\label{im:istogramma}}
\end{figure}

Previous research on J1023 has revealed that the short-term variations observed during the source's disc state, particularly the switches between the well-established low and high modes, demonstrate a notable bimodal distribution of the source counts, highlighted in particular in the X-ray light curve(\citealt{Shahbaz2015}; \citealt{Linares2014}; \citealt{Bogdanov2015}; \citealt{CotiZelati2018}; \citealt{BaglioZelati2023}). Since the observation in the optical band was not covered by an X-ray observation, we do not know when the different high and low modes occurred. We tried anyway to search for evidence of the same behaviour from the observations in the optical band extracting a histogram of the flux densities. This histogram was constructed by segmenting the observed flux, taken from every single spectrum, into intervals of 0.025 mJy. The resulting histogram is presented in Fig. \ref{im:istogramma}. This type of analysis did not reveal any clear sign of bi-modality in the optical flux. Searches for evidence of mode-switching in the optical/NIR light curve of J1023 have been carried out in the past, leading to both detections (\citealt{Shahbaz2015}) and non-detection (\citealt{BaglioVincetelli2019}. Therefore, considering also both the limited orbital coverage of our data and the $\sim 50$ s temporal sampling (which could prevent the detection of particularly short low mode episodes) the non-detection of mode switches is not unexpected.  

\noindent

\subsection{Equivalent width}
\begin{figure*}[h!]
\centering
{\includegraphics[width=.31\textwidth]{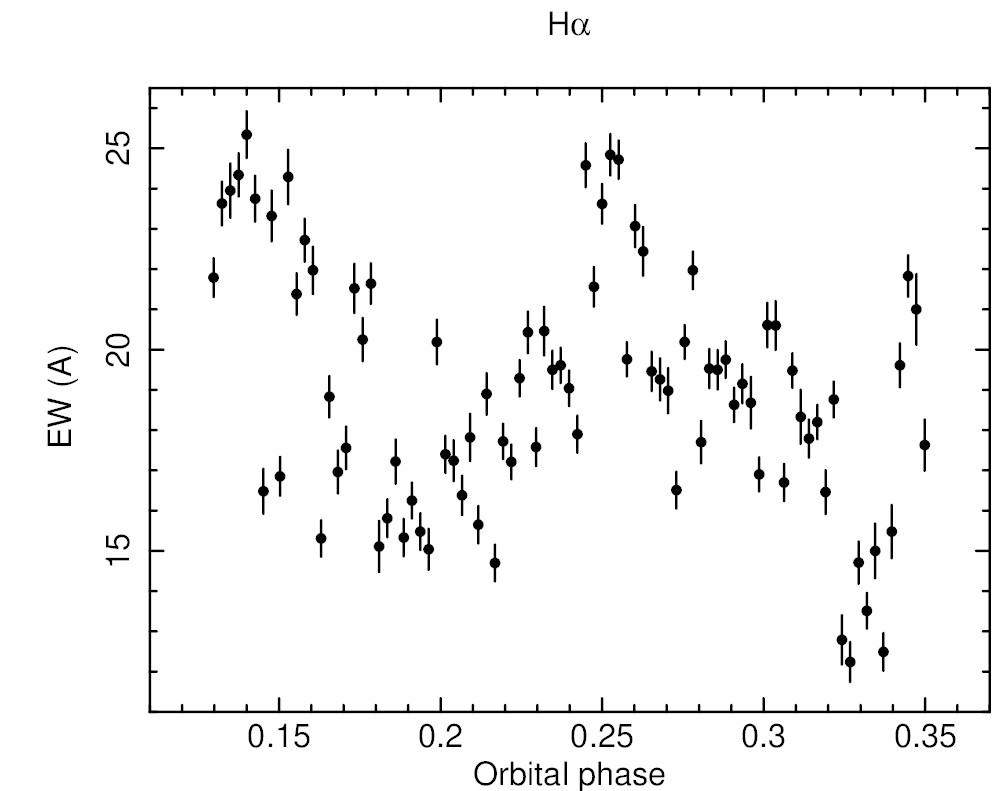}} \quad
{\includegraphics[width=.31\textwidth]{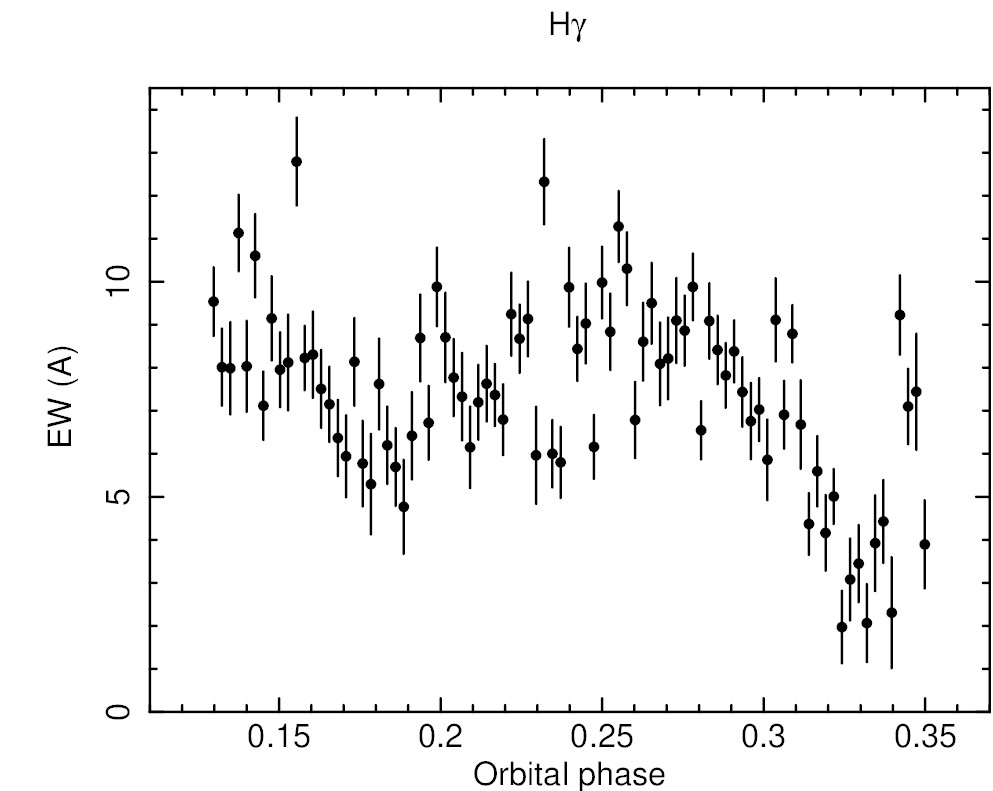}} \quad
{\includegraphics[width=.31\textwidth]{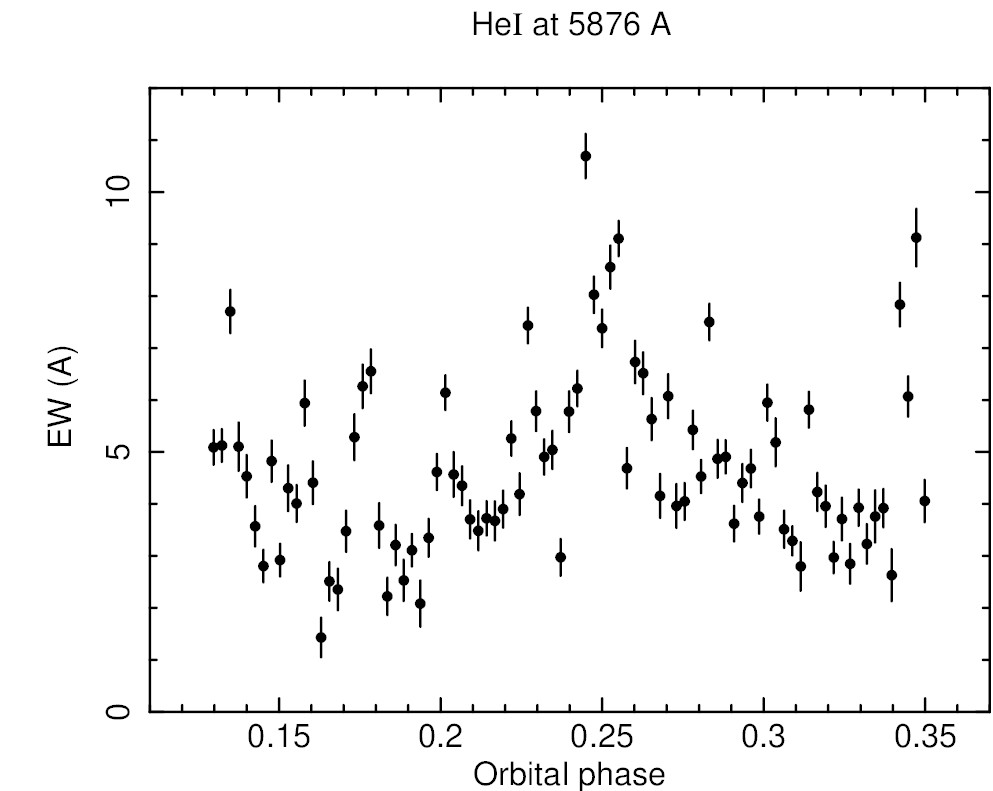}} \\
{\includegraphics[width=.31\textwidth]{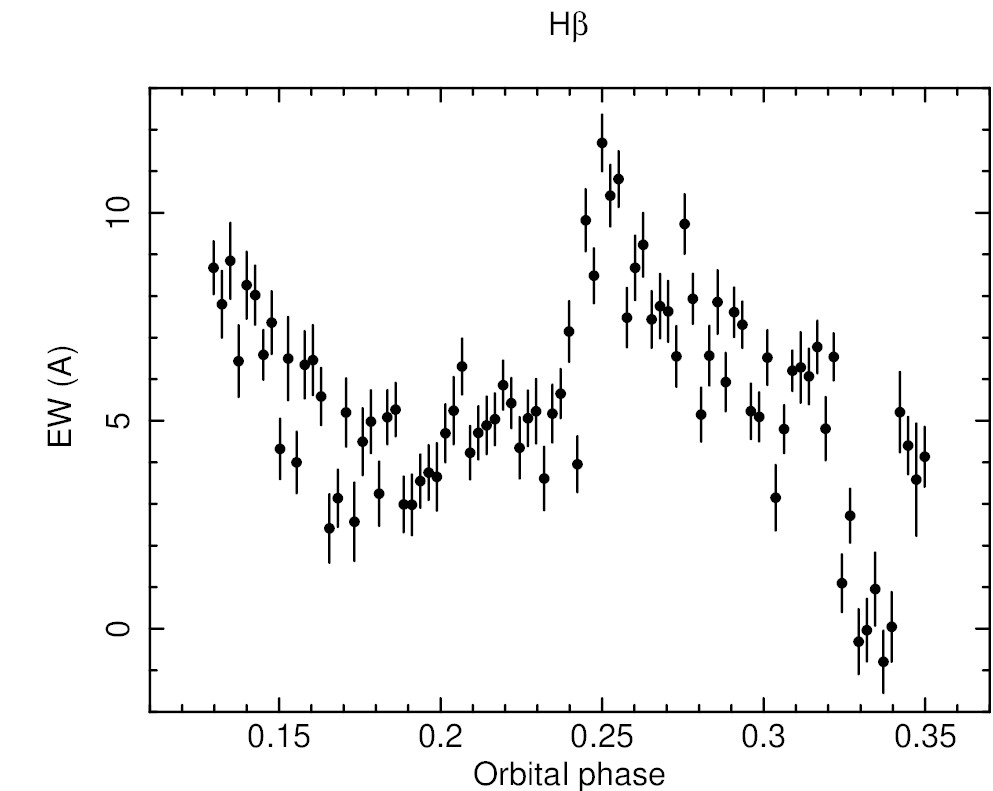}}\quad
{\includegraphics[width=.31\textwidth]{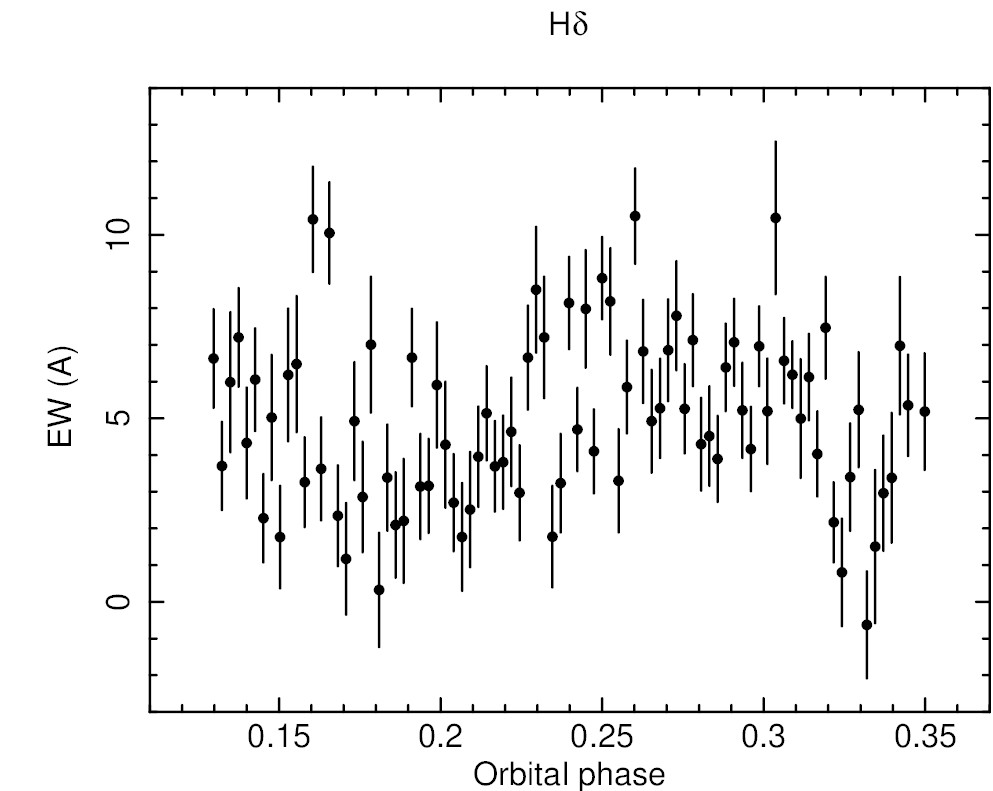}} \quad
{\includegraphics[width=.31\textwidth]{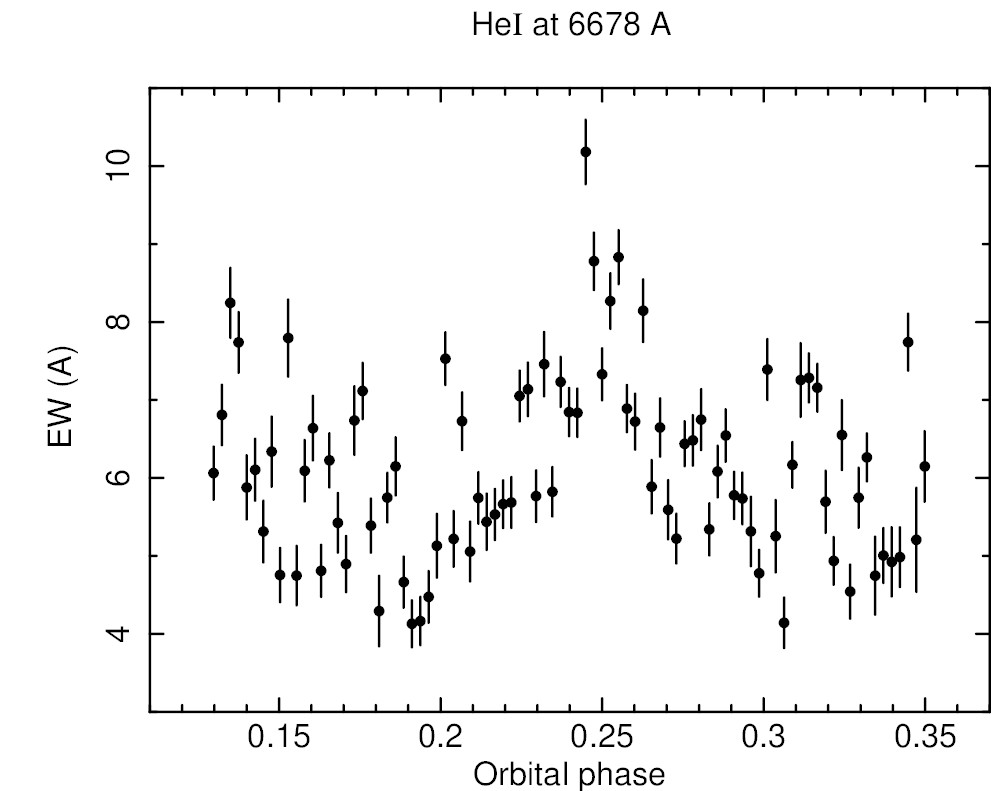}}
\caption{EW values over time for the H{\fontsize{3mm}{4mm}\selectfont $\alpha$}, H{\fontsize{3mm}{4mm}\selectfont $\beta$}, H{\fontsize{3mm}{4mm}\selectfont $\gamma$}, H{\fontsize{3mm}{4mm}\selectfont $\delta$}, He{\fontsize{3mm}{4mm}\selectfont I} at 5876 $\AA$, and He{\fontsize{3mm}{4mm}\selectfont I} at 6678 $\AA$ emission lines.}
\label{im:EWtutte}
\end{figure*}

A first indication of the intensity variability in the observed source emission was obtained by studying the equivalent width (EW)\footnote{For the study presented here, the EW measurement was made through the light task under the \texttt{molly} package, after normalising the optical continuum to the unit value.} of the various emission lines. A plot showing the Nvariability of the H{\fontsize{3mm}{4mm}\selectfont $\alpha$} line EW is shown in the top left of Fig. \ref{im:EWtutte}. As can be seen from the image, there are three clear relative maxima; namely, at orbital phases 0.14, 0.25, and 0.35. Interestingly, there is no evidence for flaring activity at the same time in the continuum emission (Fig. \ref{im:bandaV}). This behaviour is also found in the various lines, albeit it is less obvious in the weaker ones (H{\fontsize{3mm}{4mm}\selectfont $\delta$} or H{\fontsize{3mm}{4mm}\selectfont $\gamma$}, as is shown in Fig. \ref{im:EWtutte}).

\noindent

\subsection{Full width at half maximum}
\begin{figure}[h!]
\centering
    {\includegraphics[scale=0.35]{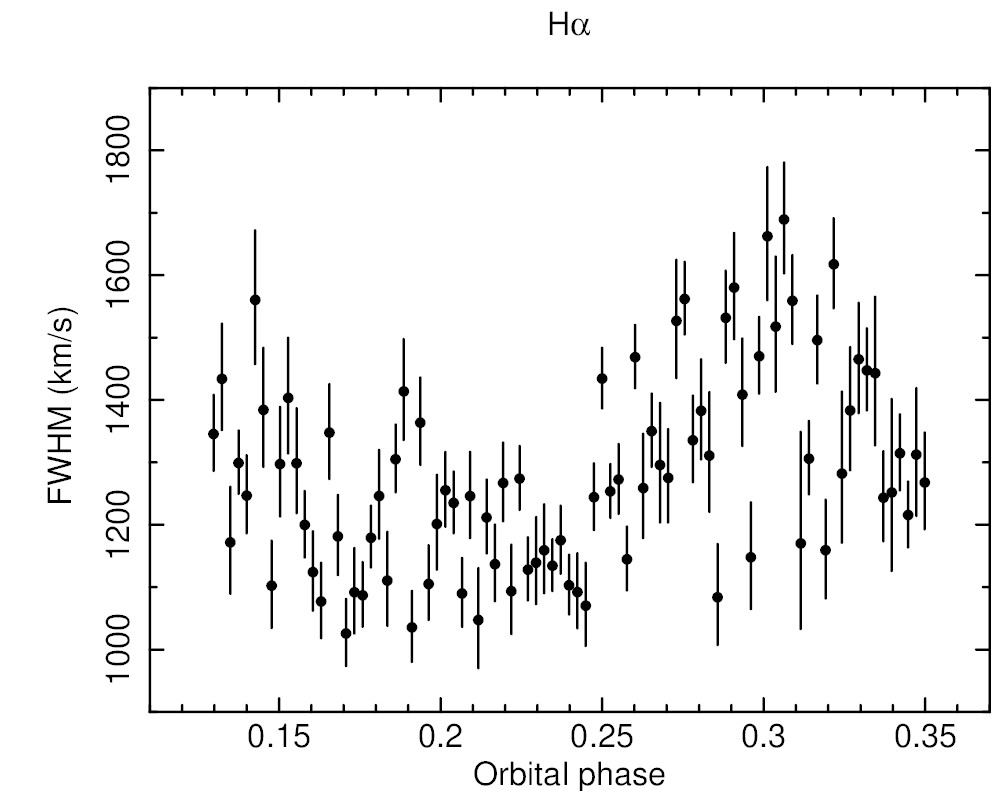}} \\
    {\includegraphics[scale=0.35]{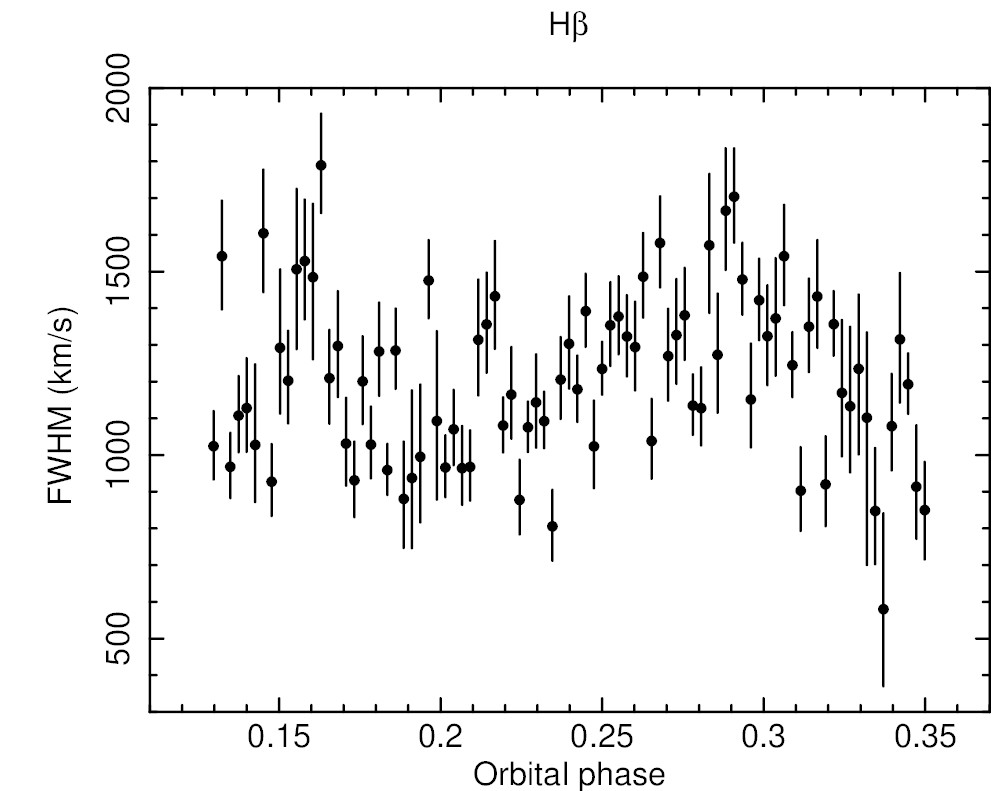}}\\
    {\includegraphics[scale=0.35]{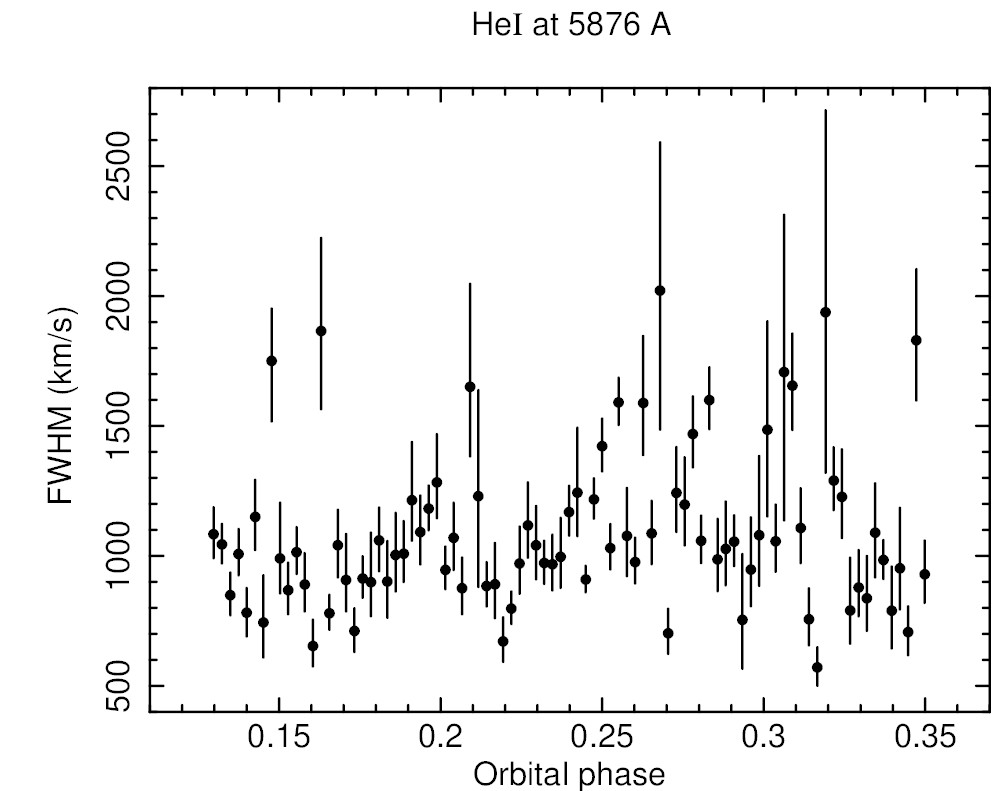}}
\caption{Three plots showing, respectively, the trend for H{\fontsize{3mm}{4mm}\selectfont $\alpha$}, H{\fontsize{3mm}{4mm}\selectfont $\beta$}, and He{\fontsize{3mm}{4mm}\selectfont I} at 5876 $\AA$ emission line of the FWHM over the orbital phase.}
\label{im:FWHMtutte}
\end{figure}

Another significant parameter for studying the line variability is the full width at half maximum (FWHM). For this study, we fitted the H{\fontsize{3mm}{4mm}\selectfont $\alpha$}, H{\fontsize{3mm}{4mm}\selectfont $\beta$}, and He{\fontsize{3mm}{4mm}\selectfont I} at 5876 $\AA$ lines (which are the most prominent emission lines for all 87 spectra), together with their underlying continuum normalised to the unity, with a Gaussian + constant model. The fit was performed on the wings of each emission line; that is, after masking the central part that is dominated by the double-horned profile, due to the presence of the accretion disc. In Fig. \ref{im:FWHMtutte}, the behaviour of the FWHM for the three emission lines is shown. As can be seen from this figure, the disc velocity behaviour does not seem to correlate with the trends seen from the EW values and the behaviour of the flux of the continuum emission (Fig. \ref{im:EWtutte} and \ref{im:bandaV}), although it does exhibit considerable variation (about 50 \% on short timescales). 

\noindent

\subsection{Emission line variability} \label{par::Emissionline}
\begin{figure}[h!]
\centering
    {\includegraphics[scale=0.35]{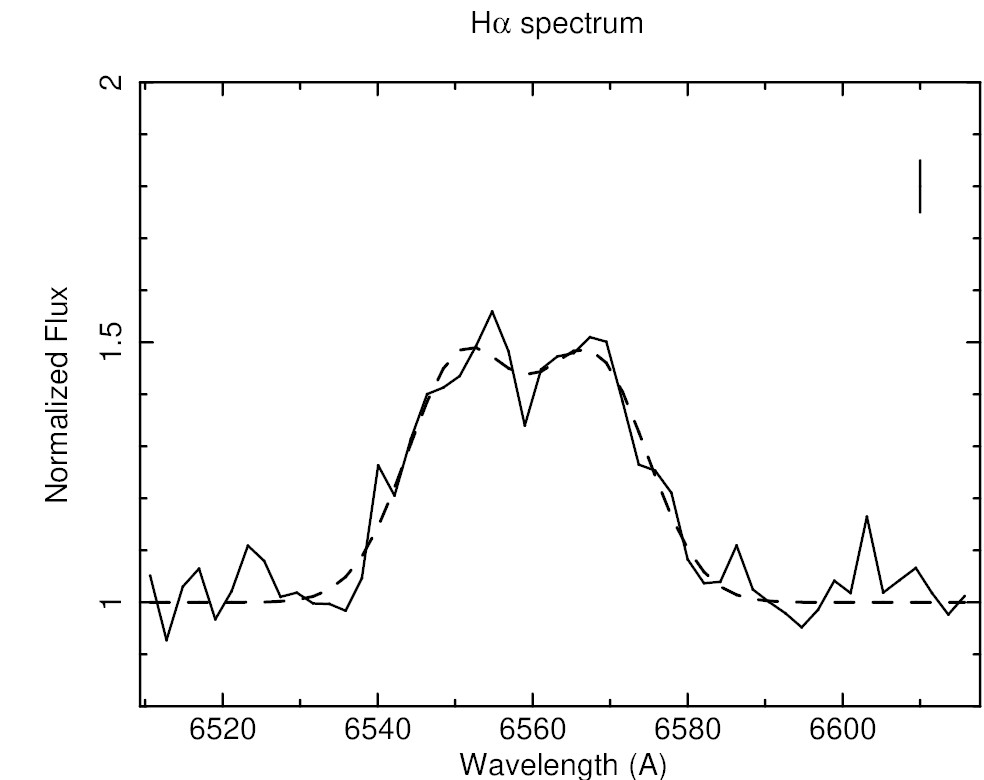}}\\
   {\includegraphics[scale=0.35]{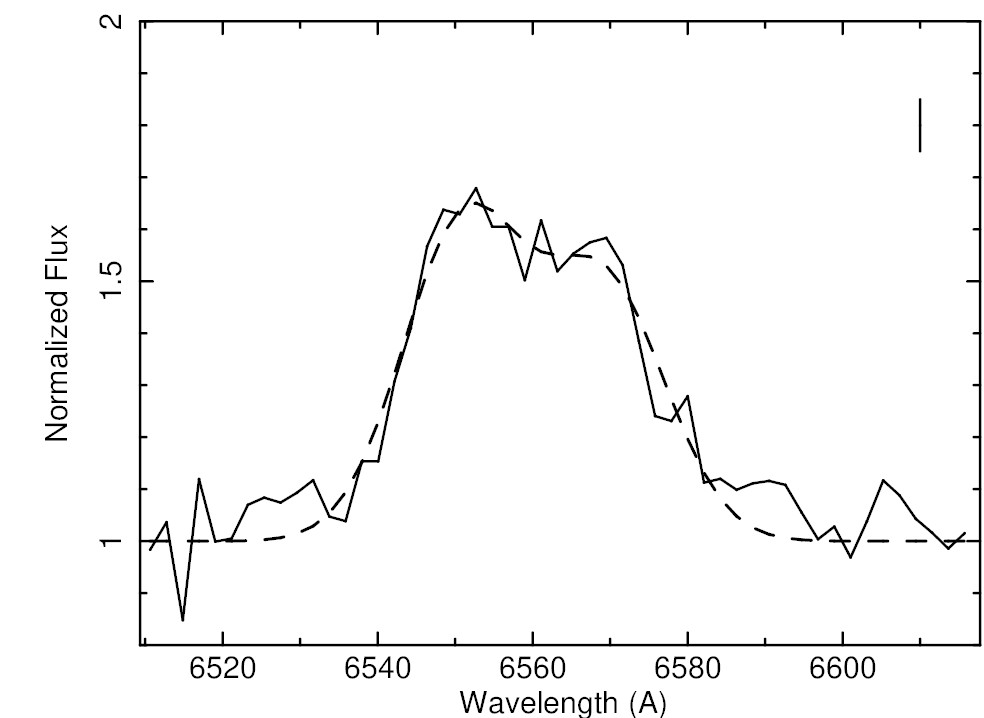}}\\
   {\includegraphics[scale=0.35]{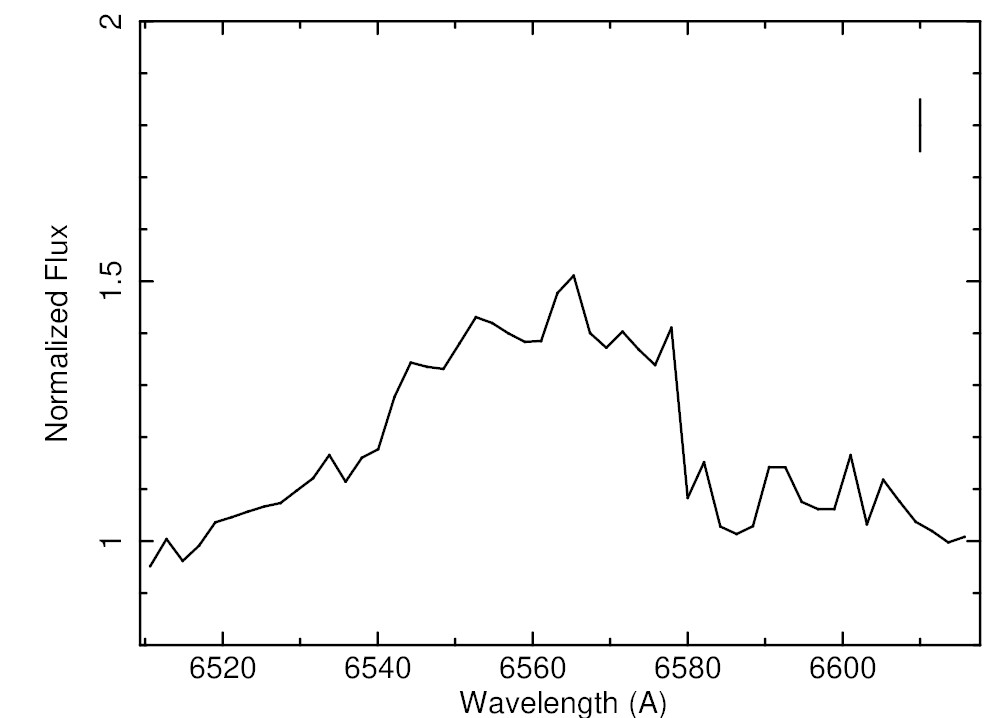}}
\caption{Three normalised spectra, with the vertical line representing the typical error, centred on the H{\fontsize{3mm}{4mm}\selectfont $\alpha$} emission line. For the first spectrum, the double Gaussian fit converges statistically within 3$\sigma$. For the second one, the fit converges statistically just below 3$\sigma$. For the last spectrum, the fit does not converge.}
\label{im:Doppipicchi}
\end{figure}

\begin{figure}[h!]
\centering
    {\includegraphics[scale=0.35]{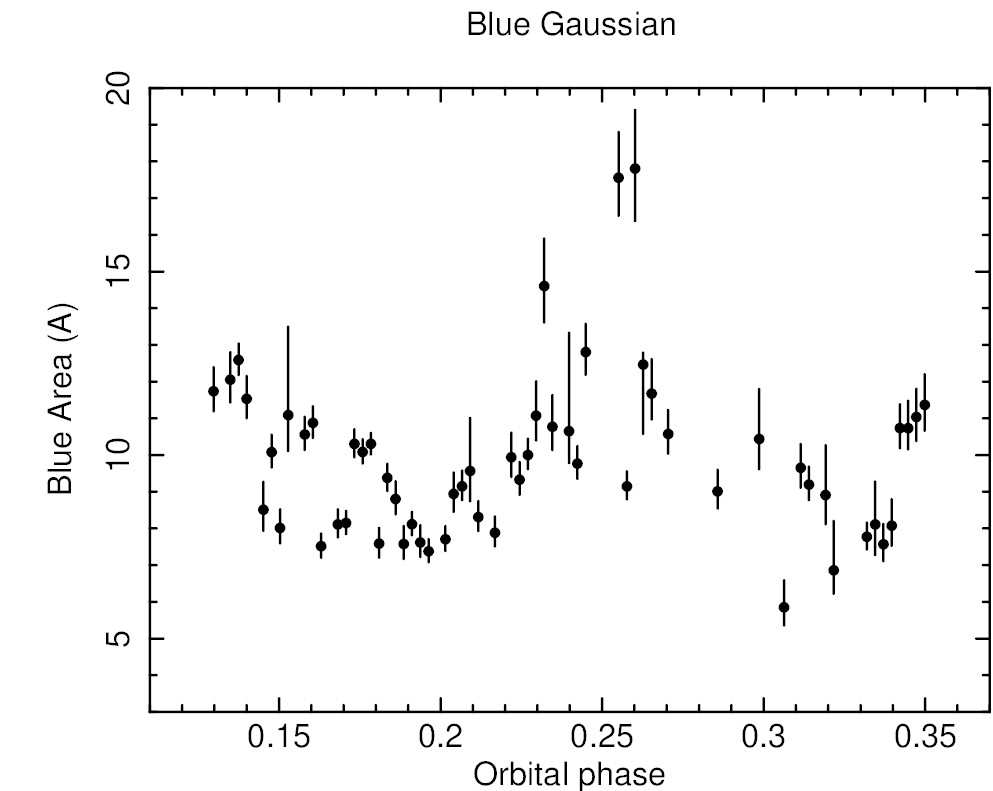}} \\
    {\includegraphics[scale=0.35]{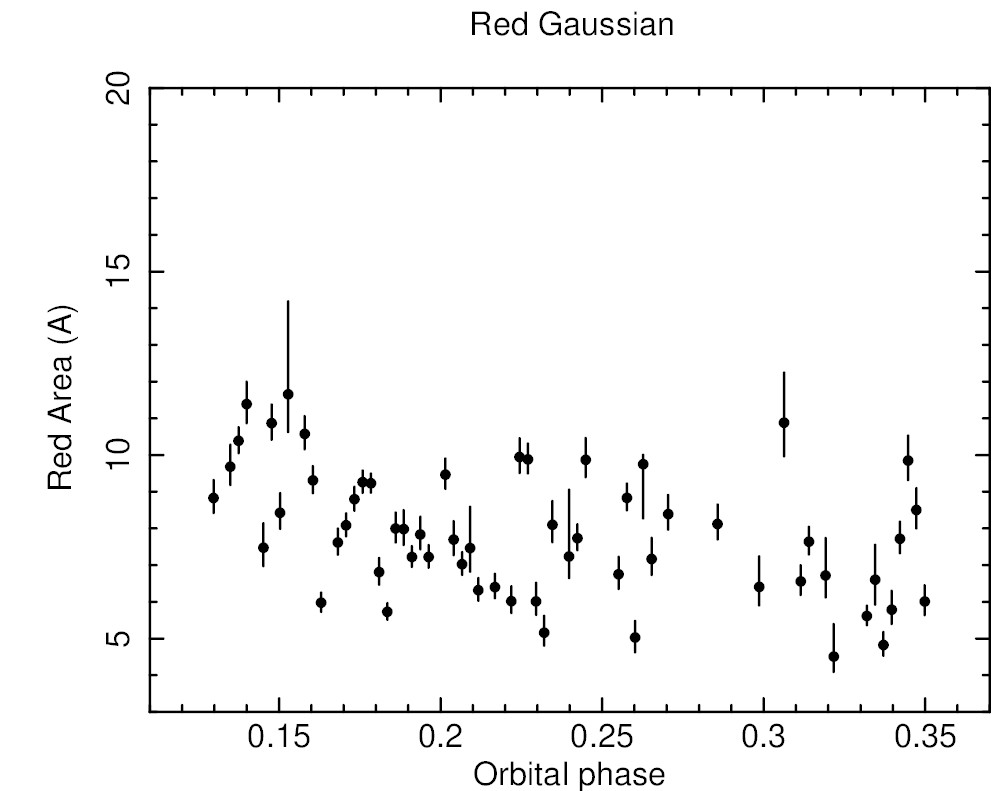}}\\
   {\includegraphics[scale=0.35]{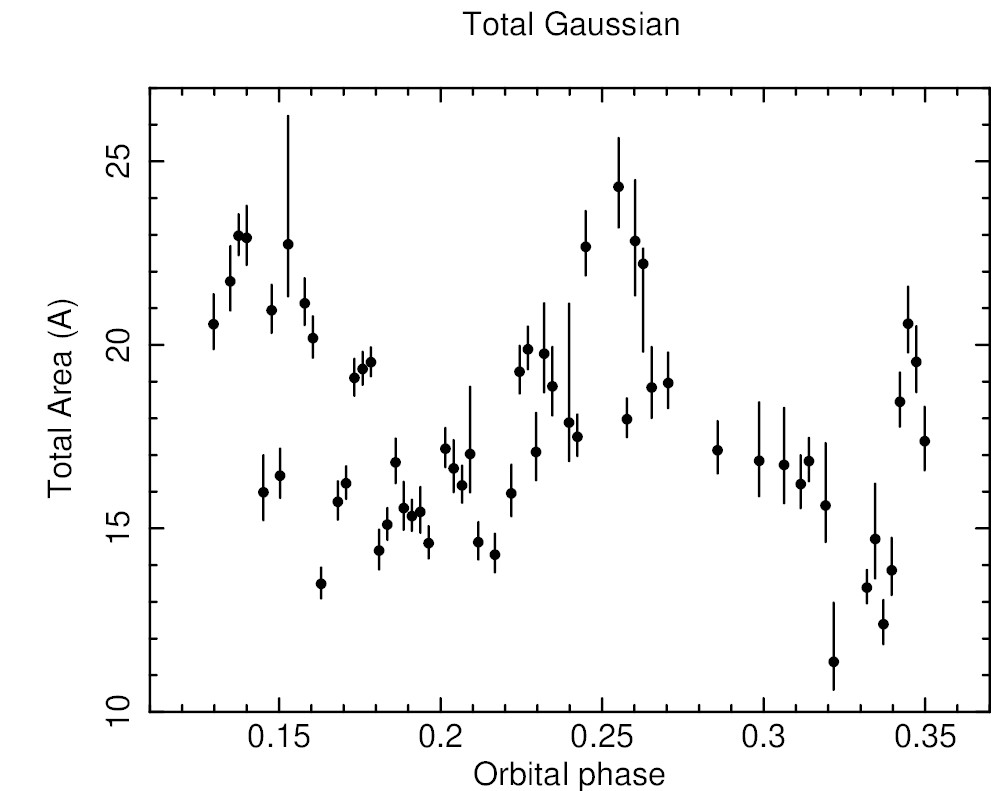}}
\caption{Three plots showing the trend of the area (A) over the orbital phase. The first two plots, show the trend of the area of the blue Gaussian and red Gaussian for the H{\fontsize{3mm}{4mm}\selectfont $\alpha$} emission line. The last one shows the sum of the areas of the blue and red Gaussians.}
\label{im:GaussiansTUTTE}
\end{figure}

Since the system is in its disc state, in order to better investigate the nature of the observed variations in line intensity, we tried to model the continuum and the expected double-horned line profile using a constant plus a double Gaussian. This analysis was specifically conducted on the most prominent line, the H{\fontsize{3mm}{4mm}\selectfont $\alpha$} (Tab. \ref{tab:appendice} in the appendix). This approach allows for the extraction of the characteristics associated with each component of the line and, potentially, the disentanglement of changes in their relative contributions to the overall line intensity over time. If the disc becomes homogeneously more luminous, it can reasonably be expected that the overall optical spectrum (continuum and emission lines) will increase in intensity. On the other hand, for a geometrically localised episode of disc variability (e.g. a flare originating from a specific disc region), one could expect a different behaviour. In such cases, the effects of a geometrically localised flare on the shape of the peak might affect only one of the two components; that is, the blue component if the flare occurs in the region where the matter in the disc is moving toward the observer, and the red component if it occurs in the region where the matter in the disc is moving away from the observer. \\
We first applied this model to the average spectrum. The fit was performed in the wavelength range (6510 -- 6618 \AA) corresponding to 51 data points. Two constraints were applied during the constant + two-Gaussian fitting process. First, the constant was frozen to 1 (since the continuum was normalised). The second constraint was to require equal FWHM values for the two different Gaussians. Every fit therefore had five free parameters (two normalisations, two centroids, and one FWHM) and 46 degrees of freedom (DOFs). These constraints are in line with expectations for a truncated disc emission and prove particularly effective for spectra in which one component (either the blue or the red) dominates over the other. We obtained a robust fit ($\tilde{\chi}^2$=18.13, DOF=46, P\footnote{Null hypothesis probability.}=0.99993). From an F-test we obtained that this model is preferred over the constant $+$ single Gaussian one by more than $3\sigma$. Having successfully tested the model on the average spectrum, we then repeated the same analysis for all 87 available spectra. When carried out on the single spectra, we find that for the majority of the spectra (59 out of 87, $\sim$ 70\%), the fit converged (Fig. \ref{im:Doppipicchi}) and the H{\fontsize{3mm}{4mm}\selectfont $\alpha$} line profile could be adequately modelled by the procedure described above, with $P$-values $>$ 1\% in about 90\% of cases. The temporal trends of these fits are presented in Fig. \ref{im:GaussiansTUTTE} (the first two plots). As is depicted in Fig. \ref{im:GaussiansTUTTE} (the first two plots), the behaviours of the blue and red components' values exhibit similarities at the start and end of the time series. However, during the mid-times, their behaviours appear markedly distinct. Specifically, the blue component exhibits significantly elevated values around phase 0.25, as was previously observed in the EW plot (see Fig. \ref{im:EWtutte}), while the red component maintains a relatively constant trend.
The reliability of the double Gaussian analysis is confirmed by the trend of the combined areas of the blue and red components over time, as is illustrated in Fig. \ref{im:GaussiansTUTTE} (the last plot). This trend closely mirrors that of the EW, which was expected since the EW serves as a proxy for line intensity.\\
For about 30\% of the spectra, the applied fit did not converge; this may be related to the fact that the H{\fontsize{3mm}{4mm}\selectfont $\alpha$} line displayed a profile with three or more peaks (Fig. \ref{im:Doppipicchi}). The presence of a third component may be due to the presence of some anisotropy; for example, the so-called hotspot (the region where the gas stream from the companion star impacts the accretion disc) or the shock front between the incoming matter from the companion and the NS pulsar wind (if present), or the companion star. The behaviour of a third peak is challenging to comprehend due to the limited observational range. Investigating how this peak changes over time would be of interest, and a comprehensive understanding of its behaviour would require observations spanning at least an entire orbital period, particularly to discern the effects related to the geometric position of the system. Another issue arises from the possible presence of multiple peaks, potentially more than three, the nature of which is difficult to interpret. Multiple peaks could be attributed to system inhomogeneities of uncertain origin or simply to noisy data. Further examination of these unusual spectra will be pursued in future work. Additionally, conducting these observations over an entire orbital period will provide valuable insights.

\noindent

\noindent

\section{Discussion}\label{Sec::Discussion}
\begin{figure}[h!]
\centering
\includegraphics[width=1\linewidth]{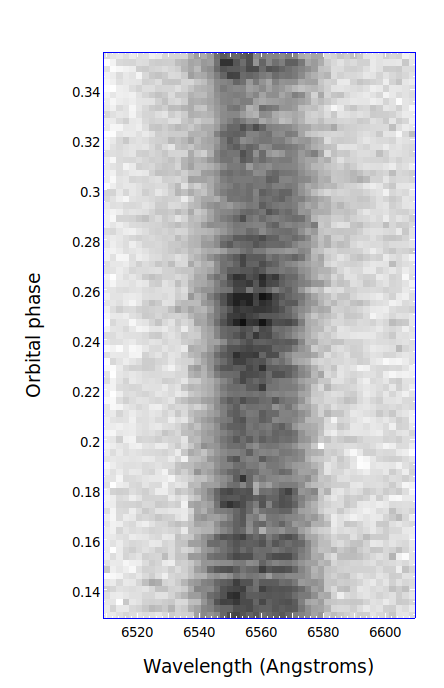}
\caption{Trailed spectrogram for the H{\fontsize{3mm}{4mm}\selectfont $\alpha$} emission line.
\label{im:Trailed_Ha}}
\end{figure}

The results presented in the previous section indicate that the optical spectrum of J1023 exhibits rapid variability on timescales of the order of minutes. Although the limited coverage of the orbital period prevents us from drawing definitive conclusions, we can state that our data do not show clear evidence of 'high-low mode' variability (Fig. \ref{im:istogramma}). On the other hand, this variability involves the continuum, the intensity, and the width of the spectral lines. Previous phase-resolved optical studies of PSR J1023 carried out since its transition to the disc state (e.g. \citealt{CotiZelati2014}) provided evidence for an optical light curve clearly modulated at the orbital period and dominated by the emission of an irradiated companion star (i.e. well modelled with a sinusoidal-like modulation with a single maximum at phase 0.5, when the companion shows its inner, irradiated face to the observer). As can be seen in Fig. \ref{im:bandaV}, the flux of the continuum does not clearly present this modulation. However, more recent optical observations of PSR J1023 suggest that such a modulation is now less evident with respect to previous studies. As an example, Fig. 3 from \citetalias{BaglioZelati2023} shows the optical light curve (REM data) of PSR J1023 observed on June 26-27, 2021 (i.e. just a few days after the GTC spectra presented in our paper). By comparing Fig. 3 of \citetalias{BaglioZelati2023} with Fig. 3 of \cite{CotiZelati2014}, it is clear that the evidence for sinusoidal-like optical modulation is definitely less evident (if not missing). On the other hand, the optical variability observed by \citetalias{BaglioZelati2023} is similar to what was observed for the optical continuum (Fig. \ref{im:bandaV}), even over a shorter orbital range. This might indicate that the disc is increasing its relative contribution to the total emission of the system at optical frequencies, gradually becoming dominant with respect to the irradiated companion component. A similar behaviour has been observed before in the quiescent accreting millisecond X-ray pulsar XTE J1814-338 (\citealt{Baglio_J1814_2013}). To thoroughly study the possible correlation between the behaviours of continuum V-band spectra, EW, and FWHM, the values were compared through Pearson correlation coefficient and Spearman’s rank correlation coefficient tests (Table \ref{tab:correlazioni}). As can be seen from the values, there does not seem to be an obvious correlation between the trend of the two parameters for H{\fontsize{3mm}{4mm}\selectfont $\alpha$} and H{\fontsize{3mm}{4mm}\selectfont $\beta$} emission lines and the continuum. In relation to the EW trend alongside the flux light curve, an inverse correlation is typically expected; in other words, as the continuum flux increases, the line becomes 'drowned out', resulting in a lower EW value. However, the peaks of EW are not matched by changes in the continuum. Similarly, the peak of the continuum does not correspond to any significant decrease in the EW value. As was described above, the behaviour of the FWHM does not appear to be correlated with the other quantities for any of the lines, although it does exhibit considerable variation in its values.\\ 
In the context of emission line variability, the results derived from the fitting of the line profile using two Gaussian functions have unveiled intriguing patterns. As is illustrated in Fig. \ref{im:GaussiansTUTTE} (a) and (b), distinct behaviours are evident during mid-times at phase 0.25, while a consistent trend is observable at the beginning and towards the end of the time series. In general, these two parameters exhibit a positive correlation, with the exception of values corresponding to the EW at the time of the maximum observed around phase $\sim$ 0.25. We also derived the trailed spectrum for the H{\fontsize{3mm}{4mm}\selectfont $\alpha$} line (Fig. \ref{im:Trailed_Ha}), although the low resolution prevents us from deriving firm conclusions. The three peaks around 0.14, 0.25, and 0.35 are clearly evident, together with a hint of an overall lower contribution of the red component to the total line emission.

\noindent
A possible interpretation of these findings suggests that the observed peaks in the overall line intensity are not solely attributable to uniform brightening of the disc. In particular, for the peak around phase 0.25, it appears that a single component, the blue one, is dominant, suggesting the presence of some anisotropy within the disc. It is evident that the blue component significantly influences the observed results. As proposed by \cite{Campana2019}, it is possible that this short-term variability is due to magnetic reconnection within the disc. This model suggests that interactions between the strong magnetic fields, generated and amplified by the differential rotation of the disc, and the accreting matter from the disc can lead to reconnections of the field lines, manifesting as intermittent 'flares' in the overall emission of the source. According to this scenario, these flares are anticipated to occur sporadically and without any discernible pattern within the inner disc region. However, alternative possibilities exist. Anisotropies in the disc structure may imply the presence of persistent bright spots that would align with the observer’s line of sight, due to the orbital motion of the system, causing periodic increases in brightness. The observational data, which covers only around 22\% of the orbital period, does not provide sufficient information to definitively distinguish between these two scenarios.

\begin{table}[h!]
\centering
\begin{tabular}{ccccc}
\toprule
\begin{tabular}[c]{@{}c@{}}\textbf{Pearson}\\ {}\end{tabular} & 
\begin{tabular}[c]{@{}c@{}}\textbf{P-value}\\ {}\end{tabular} &
\begin{tabular}[c]{@{}c@{}}\textbf{Spearman}\\ {}\end{tabular} &
\begin{tabular}[c]{@{}c@{}}\textbf{P-value}\\ {}\end{tabular}\\
\midrule
\multicolumn{5}{c}{Continuum V-band spectrum over EW value}\\
\midrule
-0.21 & 0.05  &-0.24 &0.02  &H{\fontsize{3mm}{4mm}\selectfont $\alpha$}\\
0.19 & 0.08  &0.20 &0.07      &H{\fontsize{3mm}{4mm}\selectfont $\beta$}\\
\midrule
\multicolumn{5}{c}{Continuum V-band spectrum over FWHM value}\\
\midrule
0.32 &0.00  & 0.21 &0.05      &H{\fontsize{3mm}{4mm}\selectfont $\alpha$}\\
0.37  &0.00 & 0.39 &0.00       &H{\fontsize{3mm}{4mm}\selectfont $\beta$}\\
\midrule
\multicolumn{5}{c}{FWHM over EW value}\\
\midrule
0.02 &0.84  &0.04 &0.74     &H{\fontsize{3mm}{4mm}\selectfont $\alpha$}\\
0.35 &0.00 & 0.35 &0.00     &H{\fontsize{3mm}{4mm}\selectfont $\beta$}\\
\bottomrule

    \end{tabular}
    \caption{Pearson and Spearman test values and the respective P-value for correlations between continuum V-band flux, EW and FWHM of the H{\fontsize{3mm}{4mm}\selectfont $\alpha$} and H{\fontsize{3mm}{4mm}\selectfont $\beta$} emission lines.}
    \label{tab:correlazioni}
\end{table}

\noindent

\section{Conclusions}\label{Sec::Conclusion}

 We have presented high-time-resolution spectroscopic observations of the binary tMSP J1023, in the sub-luminous disc state. The source shows, like other tMSPs, flux variability on short timescales (tens of seconds) in all bands.\\ 
Here, we have shown the results of high-time-resolution spectroscopic observations of the binary tMSP PSR J1023+0038. We obtained a total of 87 optical spectra acquired over 1.1 hours of observation, covering 22\% of the binary orbital period, making this the first optical spectroscopic study carried out on a tMSP (and an LMXB in general) with such a fast temporal cadence. Our main findings are:
\begin{enumerate}
     \item On average, each single spectrum looks rather similar to those reported in the literature for this source obtained over the same wavelength range and with longer time exposures: a blue continuum, indicative of a high-temperature disc, overlaid with intense Balmer and helium series emission lines. These lines show in most cases a double-horned emission profile indicative of the presence of an accretion disc, as was expected, the source being in its disc state at the time of its observations.
     \item We found evidence for variability in the main properties of the optical spectrum of J1023 (optical continuum, EW, and FWHM of the main emission lines) over timescales of minutes. This is the first time that variability in the spectral line properties of a tMSP has been observed over such short timescales.
     \item The episodes of variability observed in the continuum, EW, and FWHM seem to be erratic and not correlated with each other, which makes the origin of such episodes unclear.
   \end{enumerate}
     
The future development of this project is to repeat this study covering the full orbital period (4.75 hours) of the source, ideally in a multi-wavelength simultaneous observational campaign to also assess a possible correlation between the variability in the emission line properties and the mode-switching phenomenon. In the first stage, this study will verify whether the significant variability episodes are still observed, whether they have counterparts in the other bands, and whether they are sporadic or periodic. Besides, obtaining a series of optical spectra with a high time cadence over the entire orbital period would allow a Doppler map to be obtained through the technique of Doppler tomography; that is, a map in velocity coordinates displaying the geometry of the system optical emission. While a single map could not provide any direct insight into an intrinsically variable disc (or more generally, a variable emitting region), it should be possible to obtain different maps for different states of the optical spectra (e.g. flaring and not flaring; \citealt{Hakala2018}).

\begin{acknowledgements}
    We thank the referee for helpful comments. This research is based on observations made with the GTC telescope, in the Spanish Observatorio del Roque de los Muchachos of the Instituto de Astrofísica de Canarias, under Director’s Discretionary Time (Pr. ID: GTC2021-181). 

FCZ is supported by a Ramón y Cajal fellowship (grant agreement RYC2021- 030888-I), Catalan grant SGR-Cat 2021 (PI: Graber) and the program Unidad de Excelencia María de Maeztu CEX2020- 001058-M. FCZ, SCa and PD’A acknowledge financial support from INAF-Fundamental research astrophysics project “Uncovering the optical beat of the fastest magnetised neutron stars” (FANS). MCB acknowledges support from the INAF-Astrofit fellowship.
\end{acknowledgements}

%
%

\bibliographystyle{aa}
\bibliography{bibliografia}

\begin{appendix}

\onecolumn
\section{Line fitting details}
\begin{longtable}{c|c|c|c|c|c|c}
\caption{\label{kstars} Results of the fits of the H$\alpha$ line profile of J1023 with a constant + double Gaussian model. For each spectrum the exposure time is 20 s while the number of DOF is 46, corresponding to 51 points minus 5 free parameters (see Sect. \ref{par::Emissionline} for details). }\label{tab:appendice}\\
\hline\hline
Spectrum No. & Start time (UT)  & Orbital phase & Centroid n.1 ($\AA$)  & Centroid n.2 ($\AA$)  & $\chi^2$ & P-value \\
\hline
\endfirsthead
\caption{continued.}\\
\hline\hline
Spectrum No. & Start time (UT)  & Orbital phase & Centroid n.1 ($\AA$)  & Centroid n.2 ($\AA$)  & $\chi^2$ & P-value \\
\hline
\endhead
\hline
\endfoot
1 & 21:22:13 & 0.130 & 6552.2$^{+1.0}_{-0.8}$ & 6571.1$_{-1.1}^{+1.2}$ & 47.46 & 0.413 \\      
  2 & 21:24:05 & 0.135 & 6551.2$_{-0.6}^{+0.7}$ & 6569.4$_{-1.0}^{+0.9}$ & 67.43 & 0.021 \\
  3 & 21:24:49 & 0.137 & 6549.8$_{-0.6}^{+0.6}$ & 6569.4$_{-0.7}^{+0.7}$ & 40.81 & 0.689 \\
  4 & 21:25:32 &  0.140 & 6549.7$_{-0.8}^{+0.8}$ & 6569.0$_{-0.8}^{+0.8}$ & 73.92 & 0.006 \\
  5 & 21:27:01 & 0.145 & 6551.7$_{-1.1}^{+1.4}$ & 6571.2$_{-1.3}^{+1.4}$ & 52.56 & 0.235 \\
  6 & 21:27:44 & 0.148 & 6549.1$_{-0.6}^{+0.6}$ & 6567.8$_{-0.6}^{+0.6}$ & 65.18 & 0.033 \\
  7 & 21:28:29 & 0.150 & 6550.4$_{-1.1}^{+1.3}$ & 6569.7$_{-1.1}^{+1.2}$ & 39.84 & 0.727 \\
  8 & 21:29:13 & 0.153 & 6549.1$_{-0.9}^{+0.8}$ & 6567.8$_{-0.8}^{+0.8}$ & 81.14 & 0.001 \\
  9 & 21:30:40 & 0.158 & 6549.3$_{-0.8}^{+0.8}$ & 6568.6$_{-0.8}^{+0.8}$ & 44.35 & 0.542 \\
  10 & 21:31:24 & 0.160 & 6550.6$_{-0.6}^{+0.6}$ & 6569.1$_{-0.6}^{+0.7}$ & 47.96 & 0.393 \\
  11 & 21:32:07 & 0.163 & 6551.4$_{-0.6}^{+0.6}$ & 6569.2$_{-0.7}^{+0.7}$ & 60.72 & 0.072 \\
  12 & 21:33:36 & 0.168 & 6550.7$_{-0.8}^{+0.8}$ & 6569.2$_{-0.9}^{+0.8}$ & 96.59 & 1.874 \\
  13 & 21:34:19 & 0.170 & 6550.8$_{-0.6}^{+0.6}$ & 6567.9$_{-0.6}^{+0.6}$ & 53.42 & 0.211 \\
  14 & 21:35:03 & 0.173 & 6550.8$_{-0.5}^{+0.5}$ & 6568.6$_{-0.5}^{+0.5}$ & 59.75 & 0.084 \\
  15 & 21:35:47 & 0.176 & 6551.3$_{-0.6}^{+0.6}$ & 6568.6$_{-0.6}^{+0.6}$ & 47.38 & 0.416 \\
  16 & 21:36:31 & 0.178 & 6550.1$_{-0.5}^{+0.5}$ & 6569.4$_{-0.5}^{+0.5}$ & 80.28 & 0.001 \\
  17 & 21:37:15 & 0.181 & 6550.5$_{-0.7}^{+0.7}$ & 6569.9$_{-0.7}^{+0.8}$ & 50.33 & 0.306 \\
  18 & 21:37:58 & 0.184 & 6553.8$_{-0.5}^{+0.5}$ & 6571.9$_{-0.8}^{+0.9}$ & 44.57 & 0.532 \\
  19 & 21:38:41 & 0.186 & 6549.9$_{-0.9}^{+0.8}$ & 6568.0$_{-0.9}^{+0.9}$ & 47.21 & 0.423 \\
  20 & 21:39:25 & 0.189 & 6549.1$_{-1.6}^{+1.4}$ & 6566.5$_{-1.7}^{+1.4}$ & 36.75 & 0.833 \\
  21 & 21:40:08 & 0.191 & 6551.2$_{-0.6}^{+0.7}$ & 6567.9$_{-0.7}^{+0.7}$ & 53.99 & 0.196 \\
  22 & 21:40:53 & 0.194 & 6551.1$_{-1.2}^{+1.1}$ & 6568.8$_{-1.3}^{+1.1}$ & 52.08 & 0.249 \\
  23 & 21:41:37 & 0.196 & 6549.3$_{-0.7}^{+0.6}$ & 6566.2$_{-0.6}^{+0.6}$ & 57.36 & 0.122 \\
  24 & 21:43:06 & 0.202 & 6548.8$_{-1.2}^{+1.0}$ & 6566.7$_{-1.0}^{+0.8}$ & 37.26 & 0.817 \\
  25 & 21:43:49 & 0.204 & 6551.0$_{-0.9}^{+1.0}$ & 6567.9$_{-1.0}^{+1.1}$ & 40.65 & 0.695 \\
  26 & 21:44:33 & 0.207 & 6551.5$_{-0.8}^{+0.8}$ & 6567.8$_{-0.9}^{+1.0}$ & 44.90 & 0.518 \\
  27 & 21:45:16 & 0.209 & 6552.6$_{-1.5}^{+4.8}$ & 6569.1$_{-1.8}^{+4.3}$ & 48.46 & 0.374 \\
  28 & 21:46:00 & 0.212 & 6552.6$_{-0.7}^{+0.7}$ & 6568.8$_{-0.9}^{+0.9}$ & 41.09 & 0.678 \\
  29 & 21:47:28 & 0.217 & 6552.3$_{-0.8}^{+0.8}$ & 6568.2$_{-0.9}^{+1.0}$ & 45.57 & 0.490 \\
  30 & 21:48:55 & 0.222 & 6554.4$_{-0.9}^{+1.2}$ & 6569.5$_{-1.3}^{+1.4}$ & 40.33 & 0.708 \\
  31 & 21:49:38 & 0.224 & 6551.3$_{-1.0}^{+0.9}$ & 6567.5$_{-1.0}^{+0.9}$ & 52.42 & 0.239 \\
  32 & 21:50:23 & 0.227 & 6550.8$_{-0.8}^{+0.8}$ & 6567.2$_{-0.8}^{+0.8}$ & 64.73 & 0.036 \\
  33 & 21:51:06 & 0.230 & 6552.8$_{-1.0}^{+2.0}$ & 6567.2$_{-1.5}^{+2.9}$ & 43.04 & 0.597 \\
  34 & 21:51:49 & 0.232 & 6554.3$_{-1.2}^{+1.8}$ & 6571.1$_{-2.4}^{+6.0}$ & 40.35 & 0.707 \\
  35 & 21:52:33 & 0.235 & 6553.1$_{-1.1}^{+2.1}$ & 6567.7$_{-1.3}^{+1.8}$ & 63.11 & 0.048 \\
  36 & 21:54:01 & 0.240 & 6553.9$_{-0.5}^{+0.5}$ & 6566.3$_{-0.8}^{+0.8}$ & 32.77 & 0.929 \\
  37 & 21:54:45 & 0.242 & 6552.8$_{-0.7}^{+0.8}$ & 6568.2$_{-0.9}^{+0.9}$ & 40.65 & 0.695 \\
  38 & 21:55:29 & 0.245 & 6552.9$_{-0.8}^{+1.0}$ & 6567.6$_{-1.0}^{+1.0}$ & 70.67 & 0.011 \\
  39 & 21:58:23 & 0.255 & 6555.2$_{-1.6}^{+2.2}$ & 6570.6$_{-2.4}^{+5.5}$ & 40.29 & 0.709 \\
  40 & 21:59:07 & 0.258 & 6550.2$_{-0.8}^{+0.8}$ & 6566.6$_{-0.8}^{+0.8}$ & 58.30 & 0.105 \\
  41 & 21:59:51 & 0.260 & 6557.5$_{-2.5}^{+2.3}$ & 6575.8$_{-4.7}^{+12.5}$ & 35.61 & 0.866 \\
  42 & 22:00:34 & 0.263 & 6558.9$_{-2.6}^{+3.3}$ & 6559.6$_{-3.7}^{+2.9}$ & 61.12 & 0.067 \\
  43 & 22:01:18 & 0.265 & 6553.6$_{-1.2}^{+1.7}$ & 6570.5$_{-2.2}^{+1.7}$ & 26.39 & 0.991 \\
  44 & 22:02:46 & 0.270 & 6552.2$_{-0.8}^{+0.8}$ & 6569.9$_{-0.9}^{+0.9}$ & 59.00 & 0.095 \\
  45 & 22:07:08 & 0.286 & 6552.0$_{-0.7}^{+0.7}$ & 6570.3$_{-0.8}^{+0.8}$ & 82.21 & 0.001 \\
  46 & 22:10:48 & 0.299 & 6554.9$_{-1.6}^{+4.6}$ & 6572.2$_{-3.3}^{+4.3}$ & 53.67 & 0.204 \\
  47 & 22:12:60 & 0.306 & 6546.5$_{-7.8}^{+4.7}$ & 6565.4$_{-4.9}^{+2.8}$ & 49.81 & 0.324 \\
  48 & 22:14:28 & 0.311 & 6553.0$_{-0.7}^{+0.7}$ & 6573.7$_{-1.0}^{+1.1}$ & 62.21 & 0.056 \\
  49 & 22:15:11 & 0.314 & 6551.3$_{-0.6}^{+0.7}$ & 6570.0$_{-0.9}^{+0.9}$ & 41.55 & 0.659 \\
  50 & 22:16:39 & 0.319 & 6550.5$_{-0.7}^{+0.6}$ & 6567.3$_{-0.8}^{+0.8}$ & 72.22 & 0.008 \\
  51 & 22:18:51 & 0.322 & 6549.1$_{-1.6}^{+3.0}$ & 6568.1$_{-2.2}^{+3.8}$ & 52.89 & 0.225 \\
  52 & 22:20:19 & 0.332 & 6549.5$_{-0.9}^{+0.9}$ & 6570.1$_{-1.2}^{+1.1}$ & 36.53 & 0.840 \\
  53 & 22:21:03 & 0.335 & 6549.5$_{-2.1}^{+1.8}$ & 6569.7$_{-3.1}^{+2.0}$ & 47.42 & 0.415 \\
  54 & 22:21:46 & 0.337 & 6550.3$_{-1.1}^{+1.2}$ & 6568.9$_{-1.7}^{+1.9}$ & 43.45 & 0.580 \\
  55 & 22:22:31 & 0.340 & 6551.7$_{-0.8}^{+0.8}$ & 6570.7$_{-1.1}^{+1.1}$ & 53.25 & 0.215 \\
  56 & 22:23:14 & 0.342 & 6550.9$_{-0.7}^{+0.7}$ & 6569.8$_{-0.9}^{+0.9}$ & 59.20 & 0.092 \\
  57 & 22:23:57 & 0.345 & 6550.3$_{-1.1}^{+1.3}$ & 6567.7$_{-1.5}^{+1.2}$ & 56.26 & 0.143 \\
  58 & 22:24:40 & 0.347 & 6549.9$_{-0.6}^{+0.6}$ & 6569.9$_{-0.7}^{+0.7}$ & 96.30 & 2.032 \\
  59 & 22:25:25 & 0.350 & 6551.9$_{-0.9}^{+1.0}$ & 6572.2$_{-1.6}^{+2.1}$ & 57.04 & 0.128 \\
\end{longtable}

\end{appendix}

\end{document}